\documentclass[sigconf,nonacm]{acmart}

\usepackage{xspace}
\usepackage{multirow}
\usepackage[table]{xcolor} 
\usepackage{booktabs}       
\usepackage{array}         

\usepackage[table]{xcolor}
\usepackage[most]{tcolorbox} 
\usepackage[normalem]{ulem} 
\usepackage{makecell}

\usepackage{tabularx}
\usepackage{pifont}
\usepackage{array}

\newcolumntype{R}{>{\raggedleft\arraybackslash}X}

\definecolor{pos}{HTML}{8AB174} 
\definecolor{neg}{HTML}{B16F6C} 
\definecolor{neu}{HTML}{A1A1A1} 

\newcommand{\hlcell}[3][100]{%
  \begingroup
  \setlength{\fboxsep}{0pt}%
  \vphantom{\rule{0pt}{1ex}}%
  \colorbox{#2!#1}{\makebox[3em][c]{\strut #3}}%
  \endgroup
}

\newcommand{\bgcell}[3][100]{%
  \begingroup
  \setlength{\fboxsep}{0pt}%
  \vphantom{\rule{0pt}{1ex}}%
  \colorbox{#2!#1}{\makebox[2em][c]{\strut #3}}%
  \endgroup
}

\definecolor{color0}{rgb}      {0.0,0.0,0.0}
\definecolor{color1}{rgb}      {0.9,0.5,0.0}
\definecolor{color2}{rgb}      {0.6,0.0,0.8}
\definecolor{color3}{rgb}      {0.0,0.4,0.5}
\definecolor{color4}{rgb}      {0.8,0.3,0.5}
\definecolor{color5}{rgb}      {0.9,0.0,0.0}
\definecolor{color6}{rgb}      {0.0,0.6,0.8}
\definecolor{color7}{rgb}      {0.5,0.4,0.0}
\definecolor{color14}{rgb}      {1.0,0.0,0.0}

\definecolor{color8}{HTML}     {417DD7}
\definecolor{color9}{HTML}     {E85C6C}
\definecolor{color10}{HTML}    {EEBA33}
\definecolor{color11}{HTML}    {65C079}
\definecolor{color12}{HTML}    {B172CF}
\definecolor{color13}{HTML}    {8AB174}

\definecolor{darkgrey-01}{HTML}    {4c4c4f}
\definecolor{darkgrey-02}{HTML}    {2a2a2e}

\newcommand{\revise}[1]{\textcolor{color0}{#1}}

\newcommand{\pquote}[2]{%
\begin{quote}\textit{``#1''}
\textsc{~-#2}\end{quote}%
}

\newcommand{\RQtwo}{How do reviewers verify inline citations, what shapes their verification practices, and how might GenAI support citation verification?\xspace}

\newcommand{\RQfourone}{What factors are associated with reviewers’ perceptions and practices of inline citation verification?\xspace}
\newcommand{\RQfourtwo}{What challenges do reviewers encounter when verifying inline citations, and how do they wish to address them?\xspace}

\usepackage{booktabs}   
\usepackage{tabularx}  
\usepackage{array}    
\usepackage{enumitem}
\usepackage{multirow}
\usepackage{threeparttable}

\AtBeginDocument{%
  }

\setcopyright{acmlicensed}
\copyrightyear{2018}
\acmYear{2018}
\acmDOI{XXXXXXX.XXXXXXX}

\begin{document}

\title{Rethinking Inline Citation Verification in Scholarly Communication}

\author{Xinrui Fang}
\email{xinrui.fang@iis-lab.org}
\affiliation{%
  \institution{IIS Lab, The University of Tokyo}
  \city{Tokyo}
  \country{Japan}
}

\author{Reese Fairchild}
\email{rfair@cs.washington.edu}
\affiliation{%
  \institution{University of Washington}
  \city{Seattle}
  \country{United States}
}

\author{Nasi Wang}
\email{nasi.wang@iis-lab.org}
\affiliation{%
  \institution{IIS Lab, The University of Tokyo}
  \city{Tokyo}
  \country{Japan}
}

\author{Anran Xu}
\email{anran.xu@riken.jp}
\affiliation{%
  \institution{RIKEN Center for Advanced Intelligence Project, RIKEN}
  \city{Tokyo}
  \country{Japan}
}

\author{Simo Hosio}
\email{simo.hosio@oulu.fi}
\affiliation{%
  \institution{Centre for Applied Computing, University of Oulu}
  \city{Oulu}
  \country{Finland}
}

\author{Sylvain Malacria}
\email{sylvain.malacria@inria.fr}
\orcid{0000-0002-5201-5875}
\affiliation{%
  \institution{Univ. Lille, Inria, CNRS, Centrale Lille, UMR 9189 CRIStAL}
  \city{Lille}
  \country{France}
}

\author{Koji Yatani}
\email{koji@iis-lab.org}
\affiliation{%
  \institution{IIS Lab, The University of Tokyo}
  \city{Tokyo}
  \country{Japan}
}

\renewcommand{\shortauthors}{Fang et al.}

\begin{abstract}
Inline citations are central to scholarly communication, yet \revise{verifying their use} is becoming increasingly challenging because of growing review pressures. 
We conducted a mixed-methods exploratory study to investigate how reviewers verify inline citations, the factors shaping their verification practices, and how they envision GenAI supporting this process.
Across interviews (n=12) and a survey (n=203) of reviewers from HCI and AI venues, we found that reviewers’ perceptions of citation importance, verification practices, and desired AI autonomy varied across citation types, reviewer characteristics, and research backgrounds.
These findings highlight the need for adaptive support tailored to different citation types and reviewer \revise{practices}, while revealing diverse preferences regarding the use of GenAI for this process.
Moreover, effective citation verification should involve collaboration among reviewers, authors, and the broader research community, rather than relying solely on individual reviewers.
  \end{abstract}



\maketitle


\section{Introduction}

\pquote{The uncritical citation of disputed data by a writer, whether it be deliberate or not, is a serious matter.}
{P. Thomasson and J. C. Stanley (1955)}

Scientific knowledge is built upon interconnected chains of evidence. 
Inline citations play a critical role in sustaining this scholarly infrastructure by connecting claims to their supporting evidence.
Researchers use inline citations to acknowledge intellectual contributions and establish priority~\cite{merton1957priorities, kaplan1965norms}, provide evidential support for their arguments~\cite{tranoy1976foundations}, justify methodological decisions~\cite{zhang2022citational}, and situate their work within broader scholarly conversations~\cite{hyland1999academic}.

Given the important role of inline citations in scholarly communication, peer review serves as the main mechanism for maintaining the integrity and appropriate use of citations. However, the rapid adoption of LLMs, the increasing volume of scientific submissions and publications~\cite{fang2025exploring}, together with the growing number of inline citations within manuscripts~\cite{jonas2025past} are placing unprecedented pressure on existing peer review practices.
In this context, verifying the correctness and integrity of inline
citations becomes an increasingly demanding task for reviewers.
In fact, recent evidence suggests that citation integrity is becoming an emerging challenge.
A collaborative analysis by Nature and Grounded AI~\cite{naddaf2026hallucinated} estimated that tens of thousands of papers may contain at least one invalid reference generated by AI.
Similarly, in rapidly growing fields such as AI,
Sakai et al.~\cite{sakai2026hallucitation} analyzed preprints from the top tier AI conference ACL, NAACL, and EMNLP published in 2024 and 2025, identifying approximately 300 out of 17,000 papers containing hallucinated citations, with the prevalence exceeding acceptance rates in some submission cycles.

Beyond the hallucination issue, distorted citations can spread misleading information to the broader public~\cite{west2021misinformation}.
For example, a short letter published in the 1980s stated that \textit{``despite widespread use of narcotic drugs in hospitals, the development of addiction is rare in medical patients with no history of addiction''}~\cite{porter1980addiction}.
This letter has been cited more than 1,700 times on Google Scholar as of September 2026, often as evidence to support the claim that opioids are not addictive. 
However, by 2017, Leung et al.~\cite{leung20171980} showed that fewer than 20\% of these citations acknowledged that the original report was limited to hospital settings and did not apply to in-home use, where much of opioid addiction occurs.

\revise{These issues expose a critical blind spot in peer review: 
authors may selectively cite supporting evidence without sufficiently engaging with the cited sources~\cite{fang2025exploring}, while reviewers may be unable to systematically verify all of these citations.} 
Meanwhile, junior reviewers often lack guidance on how to evaluate inline citations~\cite{seals2000manuscript}. 
Together, these factors leave inline citations underlying scholarly claims vulnerable to errors and selective use that may go unchecked.
There are growing concerns about how peer review should be conducted, what criteria reviewers should use to evaluate scholarly work, and how reviewers can be better supported in this process, as reflected in recent discussions in HCI~\cite{jonas2025meta-hci,duchowski2026does}. In particular, inline citations play an important role in substantiating scholarly claims, yet how they are verified during peer review remains poorly understood.
We therefore need to understand:

\textbf{RQ1:} \RQtwo

Such insights can reveal problematic gaps in citation verification and inform the design of interactive systems to support reviewers.
Thus, we interviewed 12 reviewers and found that reviewers generally considered inline citations less critical than other aspects of a paper and selectively prioritized citations. 
Building on these findings, we conducted a large-scale survey of 203 researchers with peer-review experience who published at CHI 2026 or AAAI 2026 to understand the following questions:

\textbf{RQ2.1:} \RQfourone

\textbf{RQ2.2:} \RQfourtwo

We found that reviewers’ perceived importance of, verification practices for, and desired AI autonomy regarding inline citations varied across citation types. 
These perceptions and practices were also associated with guideline awareness, perceived citation trustworthiness, and their research domain, while desired AI autonomy was associated with confidence in AI and reviewer's geographic region. 
Together, these findings reveal how inline citation verification varies across citation types and reviewer characteristics, while highlighting challenges and opportunities for better supporting reviewers. 

Concretely, our paper makes the following contributions:

(1) An interview study with peer reviewers ($n=12$) that uncovers how reviewers selectively verify inline citations, the challenges that shape their verification practices, and their perceptions of future citation verification practices.

(2) A large-scale survey of researchers ($n=203$) with peer-review experience that characterizes how the perceived importance, verification coverage, verification effort, and desired AI autonomy vary across citation types and reviewer characteristics, and identifies the challenges reviewers encounter and the support they desire during citation verification.

(3) Design implications and considerations grounded in these empirical findings that provide directions for facilitating more effective and accountable inline citation verification in scholarly peer review.


\section{Related Work}

\subsection{Inline Citation Practice and Misuse}
Inline citations \revise{play an important role in scholarly communication by contextualizing claims to supporting sources.}
Informationally, they establish knowledge priority~\cite{merton1957priorities} and trace relationships among scientific publications~\cite{garfield1955citation}. 
Rhetorically, they strengthen claims, engage with existing research, and establish authorial credibility~\cite{Gilbert1977Referencing,case2000can,hyland1999academic}.
Socially, they recognize and reward scholarly contributions, forming part of the broader reward system of science~\cite{merton1957priorities,merton1968matthew}. 
These functions also vary across contexts: individual citations may serve methodological, comparative, or topical purposes~\cite{valenzuela2015identifying}, and can be categorized into functions such as background, methods, and results~\cite{cohan2019structural}.
Together, these perspectives position citations not merely as acknowledgments, but as integral components through which scholarly claims are contextualized.

The credibility of such claims is further shaped by peer review, which serves as a key quality-control mechanism in scholarly communication~\cite{Cronin1984TheCP}. In reviewing scholarly work, reviewers may therefore need to examine not only the claims themselves but also whether the citations accompanying those claims provide appropriate and reliable support.
This task becomes particularly important with the growing use of generative AI in scholarly writing.
\revise{Authors are responsible for ensuring that the works they cite accurately support their claims and should not cite sources without understanding their content.
Nevertheless, in practice, authors may sometimes include fabricated or inaccurate citations, particularly when using GenAI to assist with manuscript preparation~\cite{zhao2026llm}.} This raises concerns about whether inline citations accurately support the claims they accompany.
\textbf{Hallucinated citations} are among the most concerning issues associated with GenAI-assisted scholarly writing~\cite{naddaf2026hallucinated}. 
Prior work has documented their prevalence across disciplines, including AI~\cite{sakai2026hallucitation,gptzero2025_iclr2026}, medicine~\cite{cheng2024exploring}, and law~\cite{dahl2024large}. 
Such hallucinations raise concerns about the reliability and integrity of scholarly communication, as hallucinated citations can undermine researchers' trust in the evidential basis of inline citations.

Beyond hallucinated citations, \textbf{distorted citations} can also undermine the integrity of scholarly communication by misrepresenting the claims or contexts of cited work~\cite{west2021misinformation}. Prior studies have documented various forms of citation distortion, including selective citation, misrepresentation of cited claims, and the amplification of claims through citation practices~\cite{greenberg2009citation}. 
As mentioned previously, Leung et al.~\cite{leung20171980} found that a widely cited 1980 letter about opioid addiction was frequently cited without acknowledging its original limitation to hospital settings. 
Citation distortion can also propagate through a cascade effect: 
once a claim is miscited, subsequent authors may reproduce the miscitation without revisiting the original source. 
In HCI, a well-known example concerns Norman's account of affordances~\cite{norman2013design}, which has been argued to diverge from Gibson's original concept~\cite{gibson1977theory}. Although Norman later published an erratum~\cite{norman08-signifier}, the correction did not prevent the continued reproduction of the misinterpretation.
Such findings suggest that citation distortion is not merely an issue of citation accuracy, but can also alter how scientific claims are interpreted and propagated.

Despite the importance of inline citations for acknowledging prior contributions and serving sociological functions in science~\cite{kaplan1965norms,Cronin1984TheCP}, \revise{authors may lack clarity about citation norms, including when and why to cite references~\cite{klitzing2019literature}}.
Moreover, citation practices are not always systematic: Fok et al.~\cite{fok2025toward} found that researchers often search for and add citations during the writing process, which may lead authors to overlook relevant literature or inadvertently omit prior work. 
Such omissions can undermine the completeness of scholarly attribution~\cite{garfield1965can}, while practices, such as adding citations as an afterthought, may give papers an inflated appearance of scholarly credibility~\cite{mitra1970bibliographical}. 
Collectively, these challenges raise concerns about the reliability and transparency of inline citations in scholarly communication.

\subsection{Inline Citation Verification in Peer Review}

The misuse of inline citations highlights the critical role of reviewers as gatekeepers of research quality. 
Beyond the growing volume of submissions~\cite{fang2025exploring}, the density of inline citations has also increased substantially. 
For example, the average number of citations per CHI paper increased from approximately 15 in the 1980s to more than 80 in 2024~\cite{jonas2025past}.
This growth compounds longstanding challenges in peer review, including difficulties recruiting sufficient reviewers and ensuring adequate domain expertise~\cite{mccook2006peer,Vitak2025Beyond}. 
Reviewers may consequently encounter inaccurate representations of cited work~\cite{buschek2024llmspapers}, while authors themselves may not always carefully verify their citations; 
Simkin et al.~\cite{simkin2002read}, for example, reported that 80\% of authors did not read the full text of papers they cited.
Together, these challenges make systematic inline citation verification increasingly difficult for human reviewers.

To address this issue, researchers have long envisioned computational support for citation-related tasks. 
As early as 1965, Garfield~\cite{garfield1965can} envisioned automated citation indexing to facilitate the organization and retrieval of scientific references. 
However, automated verification of inline citations remains challenging because their interpretation depends on the surrounding citation context and may vary across individual citation markers~\cite{jiang2025ensembling}.
Recent advances in generative AI have opened new possibilities for this semantically demanding task. 
\revise{Although current GenAI tools can help detect hallucinated citations, they still lack the capability to verify claim–source alignment~\cite{wu2025automated}.}
Press et al.~\cite{press2024citeme}, for example, proposed an LLM-based agent for identifying papers referenced in scientific text.
However, the agent achieved only 35.3\% accuracy, substantially lower than the 69.7\% achieved by human participants, highlighting the limitations of current LLMs for reliable citation attribution.

Beyond citation-specific tasks, researchers have explored how AI can augment reviewers more broadly. 
Chen et al.~\cite{Chen2025Envisioning} found that ChatGPT-assisted reviewing reduced reviewers' cognitive load, but did not significantly reduce review time or improve review quality, suggesting that AI may be more effective as a complement to human expertise than as a replacement. 
At the same time, AI assistance introduces potential risks. Russo et al.~\cite{Giuseppe2025The} showed that AI-assisted reviewing can influence paper scores and acceptance decisions, potentially shaping future research directions through biases embedded in AI systems. 
HCI researchers have therefore explored systems that scaffold, rather than replace, human reviewers. 
For example, ReviewFlow~\cite{sun2024reviewflow} provides AI-supported scaffolding for the peer-review process, helping reviewers identify and organize strengths and weaknesses in submissions.
Similarly, MetaWriter~\cite{Sun2024MetaWriter} supports reviewers in synthesizing peer reviews and authoring meta-reviews.
However, existing systems primarily support the overall reviewing workflow rather than the specific task of verifying whether claims are adequately supported by their inline citations.

Taken together, prior work reveals a gap between the growing need for inline citation verification and the capabilities of existing approaches.
Human reviewers face increasing verification burdens, while automated approaches remain insufficiently reliable. 
Although AI-assisted peer-review systems can support reviewers more broadly, they do not specifically address the evidential relationship between claims and their cited sources.
This gap motivates us to investigate how reviewers envision collaborating with GenAI to support inline citation verification during peer review.

\section{Motivating the Study: Review Guidelines}
\label{section: review guideline}
\begin{table*}[t]
\small
\centering
\caption{Semantic themes of inline citation review guidelines with representative quotes. 
\bgcell[100]{pos}{$\bullet$} indicates that the theme is mentioned in the venue's guidelines; \bgcell[100]{neg}{$\circ$} indicates that the theme is not mentioned.}
\label{tab:review_guideline}
\begin{tabularx}{\linewidth}{p{2cm} p{9cm} cccc cccc}
\toprule
\multirow{2}{*}{Themes} &
\multirow{2}{*}{Examples} &
\multicolumn{2}{c}{HCI} &
\multicolumn{4}{c}{AI} \\
\cmidrule(lr){3-4}\cmidrule(lr){5-8}
&
&
CHI  & CSCW &
AAAI & ICLR & NeurIPS & ICML \\
\midrule
Research \newline Positioning & ``Is it clear how this work differs from previous contributions, with relevant citations provided?'' \hfill 
\textit{--- NeurIPS 2026} & \bgcell[100]{pos}{$\bullet$} & \bgcell[100]{pos}{$\bullet$} & \bgcell[100]{pos}{$\bullet$} & \bgcell[100]{neg}{$\circ$} & \bgcell[100]{pos}{$\bullet$} & \bgcell[100]{pos}{$\bullet$}\\ \addlinespace
\hline
\noalign{\vskip 3pt}
Review \newline Coverage & ``Read the paper: It’s important to carefully read through the entire paper and to look up any related work and citations that will help you comprehensively evaluate it.'' \hfill 
\textit{--- ICLR 2026} & \bgcell[100]{neg}{$\circ$} & \bgcell[100]{neg}{$\circ$} & \bgcell[100]{neg}{$\circ$} & \bgcell[100]{pos}{$\bullet$} & \bgcell[100]{neg}{$\circ$} & \bgcell[100]{neg}{$\circ$}\\ \addlinespace
\hline
\noalign{\vskip 3pt}
Citation \newline Omission \newline Tolerance & ``If some relevant prior works are missed, then think about whether or not including them would change the conclusions of the paper. Some omissions may be considered minor issues that are easily fixed.'' \hfill \textit{--- ICML 2026} & \bgcell[100]{neg}{$\circ$} & \bgcell[100]{neg}{$\circ$} & \bgcell[100]{neg}{$\circ$} & \bgcell[100]{pos}{$\bullet$} & \bgcell[100]{neg}{$\circ$} & \bgcell[100]{pos}{$\bullet$}\\  
\bottomrule
\end{tabularx}
\end{table*}

Reviewers’ practices are shaped not only by their own experience but also by umbrella review guidelines. 
Therefore, understanding how reviewers verify inline citations also requires examining how review guidelines may instruct or encourage reviewers to assess them.
Thus, we first conducted a preliminary study to examine how top-tier HCI and AI venues address inline citation verification in their reviewer guidelines and whether these guidelines provide actionable guidance for this practice.

We selected venues that represent major publication outlets across both the HCI and AI communities.
Specifically, we selected leading HCI venues (CHI, UIST, CSCW, and IMWUT) and AI venues (AAAI, ICLR, ICML, and NeurIPS) of the year 2026. 
\revise{As UIST, CSCW and IMWUT do not make their reviewer guidelines publicly accessible, but instead share them internally among reviewers,}
\revise{we contacted their organizing committees to request access to review guidelines.}
Among these venues, we were granted access to the reviewer guidelines of CSCW, while UIST and IMWUT organizing committees did not answer our requests.
We then searched the review guidelines using the following keywords, \revise{which we identified in the guidelines as being related to inline citations} (\textit{reference}, \textit{refer}, \textit{cite}, \textit{citation}, \textit{related work}, \textit{arXiv}, \textit{preprint}, \textit{technical report}, and \textit{omission}) and manually coded instructions related to inline citations. 
Two authors independently coded the guidelines using an inductive coding process~\cite{thomas2006general}. 
They subsequently discussed all discrepancies and reached consensus on the final coding results. 
The coding dimensions capturing how review guidelines address inline citations are summarized in Table~\ref{tab:review_guideline}.

\revise{Our examination suggests that these reviewer guidelines provide only limited actionable guidance on inline citation verification. 
First, regarding \textbf{what reviewers should verify}, most venues vaguely emphasize evaluating the \textit{Research Positioning} of a submission within the existing literature, but none specify how reviewers should assess whether individual inline citations are appropriately positioned or accurately support the claims they accompany.
Second, regarding \textbf{how extensively reviewers should verify inline citations}, guidance on \textit{Review Coverage} is largely absent, with the exception of ICLR 2026 that explicitly encourages reviewers to look up related work and citations, without specifying how many or which inline citations should be verified.
Third, regarding \textbf{how reviewers should respond to citation misuse}, few venues provide guidance on this aspect. ICLR 2026 and ICML 2026 explicitly suggest a degree of \textit{Omission Tolerance} for incomplete citation coverage. 
However, beyond citation omission, the guidelines provide little guidance on how reviewers should respond to problematic inline citations, such as citations that are misleading or do not adequately support the associated claims.}

These observations suggest that existing guidelines establish broad expectations for citation-related rigor, while leaving substantial room for reviewers to interpret what to verify, how extensively to verify it, and how to respond to citation misuse.
To understand how reviewers navigate this lack of procedural guidance in practice, we conducted an interview study (see Section \ref{Section: Interview Study}) guided by \textbf{RQ1:} \RQtwo 
\label{RQ1}

\section{Interview Study}
We first conducted a semi-structured interview study to develop an empirical understanding of the \textbf{RQ1}. 
The study was approved by the Institutional Review Board (IRB) at the university of the first author.

\label{Section: Interview Study}
\subsection{Participants}
We invited 12 researchers (9 male and 3 female) aged 22-39 years (M = 28.1) to participate in this interview study through authors' social connections and snowball sampling~\cite{noy2008sampling}. 
All participants had review experience at conferences ranked A* in the International CORE (ICORE) Conference Rankings\footnote{\url{https://www.core.edu.au/icore-portal}}.
Six participants were HCI researchers, among them, three had Associate Chair experience, and the other three had external review experience \revise{but had not held senior reviewing roles}; 
Another six participants were AI researchers, and among them, three had area chair or PC member experience (see Appendix \ref{tab:pr_info}). 

\subsection{Procedures}
At the beginning of the interview, we introduced the purpose of our study.
After participants confirmed their understanding of the research purpose and their participation, we invited them to recall the latest paper they had reviewed as a warm-up.
Following this warm-up question, we proceeded with the interview following the prepared interview script (see Appendix \ref{intervew_scripit_reviewers}).
The interview lasted around 60 minutes.
We compensated each participant approximately \$20 USD for their participation. We conducted all interviews online in English via Zoom and recorded them for transcription. We deleted the audio recordings immediately after transcription to protect participants’ privacy, then cleaned the transcripts and removed any personally identifiable information before analysis.

Two coders then analyzed the transcripts using an inductive thematic analysis approach~\cite{braun2006using, thomas2006general}.
The coders first familiarized themselves with the first six transcripts and independently and inductively coded four of them (PR1-PR4) to develop their initial codebooks. 
They then shared their codebooks and discussed divergent codes and missing quotations in each other's work. 
Based on this discussion, they developed a shared codebook.
Next, both coders re-coded the same four transcripts (PR1-PR4) using the shared codebook. 
Once major disagreements were resolved, they proceeded to code the next four transcripts (PR5-PR8) independently using the shared codebook. 
They again discussed any divergences, new codes, and relevant quotations, and iteratively refined the shared codebook.
Finally, the remaining four transcripts (PR9-PR12) were divided between the two coders, with each coding half (PR9-PR10; PR11-PR12) using the refined codebook.

During the coding process, we also regularly discussed the results with the other authors to support iteration and mitigate potential bias.
Through iterative refinement and discussion, we identified 154 codes across the dataset.
These codes were synthesized into themes, of which we report those most relevant to our research questions and contributions in the findings below. 
The detailed codebook, including the full set of themes and codes, is provided in the supplementary material.

\subsection{Findings}
\subsubsection{Inline Citation Verification Practices}
\revise{Participants selectively verified whether cited sources supported the claims they were used to substantiate, because they regarded citation verification as a lower-priority activity during peer review. 
Meanwhile, they differed in which citations they chose to verify and which they chose to skip.}

\paragraph{Selective verification rather than systematic checking}
\label{phase II: limited verification}
\revise{Although participants perceived inline citations as important, recognizing them as a \textit{``foundational mechanism for building upon prior work''} (PR6), this perceived importance did not necessarily translate into comprehensive verification.
Instead, they selectively decided which citations warranted closer inspection.}
Participants generally described inline citation verification as a lower priority activity than other aspects of peer review, such as assessing novelty: \textit{``Novelty should be the first priority in this work...[if the novelty is good],  I will ignore any of the other drawbacks in the related work.''} (PR3).
Rather than systematically checking whether cited sources supported the claim or not, participants tended to selectively check inline citations because of \textbf{time constraint}: \textit{``I don't have time to review every reference paper''} (PR3),
\textbf{interest driven}: \textit{``My main motivation for checking a reference is not to check if that reference is right... I am hoping that this is time that I am spending learning about the research. It’s not just to assess the paper.''} (PR11) and their \textbf{perceived author's responsibility} : \textit{``I see checking the literature as primarily the author's responsibility. If reviewers had to check every citation, that would place quite a heavy burden on them.''} (PR7).
Consequently, inline citation verification emerged less as a systematic review responsibility and more as an optional activity.

\paragraph{Prioritizing citations based on claim importance}
\label{phase II: importance level}
\revise{On the other hand, this selectivity was primarily driven by the \textbf{perceived \textit{``importance''}} of the claims that citations supported.}
Such as the citations used to motivate \textit{``research problem''} (PR6),  compare the \textit{``state-of-the-art methods''} (PR7), and state the research gap:
\textit{``Especially those that the authors closely compare to and use to state their research gap. I guess those are the papers I need to take a closer look at''} (PR1). In contrast, participants noted that they deprioritized inline citations in \textit{``background''} (PR4) or \textit{``related work''} (PR3), especially when the surrounding argument appeared reasonable or aligned with their prior knowledge.
For example, in technical papers, one participant explained that:
\textit{``If I see the related work from scratch is all reasonable, I won't spend time checking whether it is reliable''} (PR3).
These findings led us to reflect on the factors underlying this contradiction, as well as how reviewers form their perceptions of the importance of different types of citations.

\subsubsection{Challenges Shaping Verification Practices}
\revise{Participants' selective verification was not simply a matter of preference. 
Their ability to verify citations was constrained by both the limits of their knowledge and the time available to them. 
At the same time, although participants are expected to assess citations, they often lack clear procedural guidance and concern about the trustworthiness of the inline citations they encounter.}

\paragraph{Limits of expertise}
\label{phase II: challenges}

Participants described inline citation verification relies heavily on existing domain knowledge: \textit{``normally based on my experience''} (PR2).
However, this strategy became substantially less reliable when reviewing unfamiliar domains:
\textit{``It also depends on what kind of paper I'm reviewing... But if it is from a completely different area, I would probably struggle to check these things [inline citation misuse].
''}(PR5).
Some participants described the limitation even more strongly: 
\textit{``Unless you know a lot about this field. 90\% of the citations can't be verified.''} (PR12).
This limitation also affected what participants were willing to criticize. One participant explained that when unfamiliar with a domain, they tended to \textit{``avoided making 'strong claims or criticisms about work in areas I’m not confident about'''} (PR6).
\revise{Expertise determined not only how reviewers verified citations, but also which citations they felt qualified to challenge.}

\paragraph{Growing review burden}
\label{Interview-Growing}
\revise{Even when participants had sufficient expertise, they could not afford to verify everything.}
The increasing number of submissions and references places an increasing burden on participants to verify. 
One participant observed that although a paper might contain \textit{``a hundred citations''}, only \textit{``three to five papers''} (PR1) might be truly important to the paper.
Another participant described a shift associated with the increasing use of GenAIs: \textit{``Now [because of GenAI] there are more papers, the pressure has increased''} (PR8).

Among this rapid growth in submissions, the presence of low quality and non peer reviewed references further \textbf{burdens participants while offering limited opportunities to learn}:
\textit{``In the past, I could almost guarantee that I would learn something new when reading papers. But now there are so many papers, and the quality is not always high. More than half of my reading time may be occupied by papers that do not provide much value. If I read papers like this every day, I may not learn much. I don't want to spend that cost''} (PR8).

\paragraph{Lack of guidance}
\label{phase II: guideline}
Moreover, some participants reported that they did not know how to properly review inline citations, as they could not find detailed guidelines on how to do so:
\textit{``In those [review] guidelines, I did not find any pointers or references about this. I think the conference organizers did not explicitly tell me to pay attention to this aspect''} (PR1).
While others observed that citation correctness \textit{``should be a factor''} but \textit{``nowadays, we don't think it will be a factor''} (PR2),  since it is not directly mentioned in review criteria.
Another participant similarly reported having \textit{``no impression''} (PR8) of specific guidelines for reviewing citations.
These findings further corroborate our review guideline analysis (see Section \ref{section: review guideline}) by providing evidence from the reviewers’ perspective.

\paragraph{Declining reviewer trust}
\label{phase II: trustworthness}

Participants tended to approach inline citations with an initial presumption of trust, describing citation as \textit{``something everyone implicitly agrees on''} (PR6), unless they encountered cues that prompted them to question their validity: \textit{``But once I found it [mistake], I think my perception of the papers will change''} (PR1).
\revise{Yet, despite citation verification receiving limited procedural attention, participants increasingly encountered reasons to distrust citations.}
Several participants described citation problems as something they had not previously considered a major concern, but had recently become more attentive to as AI-generated references became more common. One participant noted: \textit{``Recently, people have been using AI to generate BibTeX files. This year's ICLR seems to have caused a lot of problems, so I'm a little more worried about it recently''} (PR4).
Similarly, one participant noted that they \textit{``weren't really concerned about this issue two years ago,''} but had observed \textit{``a lot of misuses of citations''} (PR12) over the past two years.
Participants also perceived GenAI as contributing to a decline in the quality and appropriateness of cited sources. One participant argued that \textit{``the quality of citations has decreased''} (PR8), describing cases in which authors cited a lower-quality paper even when a more authoritative or recent source made a similar claim.

\subsubsection{Perceptions of Future Citation Verification}
\revise{Participants faced the tension between the desire for assistance to improve verification and the concerns about the reliability of AI's output. 
Moreover, they expressed different levels of collaboration with AI in the future tasks.}
\paragraph{Limited confidence in AI verification}
\label{Phase II: Confidence}
Participants saw GenAI as a potential means of compensating for their verification limitations: 
\textit{``[Once GenAI is introduced to help summarize inline citations] It can be easier for them to verify the citations are cited correctly or properly''} (PR2).
In this sense, reducing the effort required for verification could enable reviewers to devote attention to citations that they currently overlook due to limited time or expertise.

However, participants were reluctant to fully rely on AI for citation verification because they remained uncertain about the reliability of AI-generated judgments. PR3 explained that they would use an LLM to summarize or analyze related work, but emphasized, \textit{``I don’t 100\% trust it because I worry about hallucinations''}.
Moreover, participants also raised concerns that AI-generated citations may reflect or introduce societal biases, emphasizing the importance of human involvement in the citation process: \textit{``You can't let society's biases and preferences determine what LLMs decide. No matter what, you have to incorporate people's preferences into the process. So from this perspective, I think people's participation is very important''} (PR8).
Thus, reviewers faced a \textbf{tension} between the potential efficiency of AI-assisted verification and concerns about the reliability of AI-generated judgments.

\paragraph{Human–AI Collaboration for Citation Verification}
\label{Phase II: Human-AI collabration}
To address the tension discussed above, participants described human–AI collaboration at two complementary levels. At the task level, they proposed different ways of dividing verification work between humans and AI. First, \textit{task-based division of labor} involved assigning different types of verification tasks to humans and AI. For example, participants suggested that GenAI could handle more objective tasks such as fact-checking, while human reviewers could focus on more subjective judgments:
\textit{``In the near future, I believe we can use local LLMs to create an assistant for reviewers. Like when we use Gemini, AI can first help us check whether our comments actually align with the manuscript and are grounded in what the paper says before we submit them''} (PR6).
Second, \textit{priority-based division of labor} involved allocating verification effort based on the perceived importance of different inline citations. As discussed in the previous section, participants recognized that different types of citations vary in importance and suggested delegating the verification of less important inline citations to GenAI, while personally verifying citations that are critical to the paper's framing:
\textit{``I want to check those papers by myself because they are very critical to the overall framing of a paper. But I think for papers that are only used for background or methodology, that would be fine''} (PR1).

At the decision level, participants emphasized that humans should retain authority over verification decisions. They viewed AI as a source of support rather than a substitute for human judgment, with final decisions remaining with human reviewers:
\textit{``I think I still need to be involved in the process. I need to make sure I am certain about the key parts. If I delegate all my work to AI, I don't think I can take responsibility as a reviewer anymore''} (PR4).
At the level of citation choices, participants emphasized that individual citation choices reflect diverse values and perspectives, suggesting that such diversity should be preserved in scholarly communication:
\textit{``I think the stronger the subjective expression of the paper, the more interested I am''} (PR12).

\subsection{Summary}
This interview study revealed substantial variation in how reviewers prioritize and verify different types of inline citations.
Reviewers described selectively verifying citations in response to challenges such as the increasing burden of peer review and the lack of review guidance. 
They also expressed varying levels of willingness to collaborate with GenAI in future verification practices.
These findings motivate a broader survey study (see Section \ref{Section: Survey Study}) into the factors shaping reviewers’ citation verification practices and perceptions, as well as the challenges the reviewers encounter and the mitigation strategies they desire.

\section{Survey Study}
\label{Section: Survey Study}

\revise{We further sought to examine, at a larger scale and across research communities, how broadly the patterns identified in the previous study were shared among reviewers.
Specifically, we focused on the following two research questions:}

\label{RQ2.1}
\textbf{RQ2.1:} \RQfourone

\label{RQ2.2}
\textbf{RQ2.2:} \RQfourtwo

To address these research questions, we conducted a large-scale survey study. Our survey consisted of 49 questions, including two attention-check questions: one in the first half of the survey and one in the second half (see Appendix \ref{appendix: survey}). This study also obtained separate approval from the IRB of the first author’s institution.

\subsection{Participants and Recruitment}

Following prior HCI meta-research~\cite{gronbak2026how}, we recruited participants by sending individualized email invitations to eligible authors. Our target participants were authors of full papers published at ACM CHI 2026 or AAAI 2026 who had prior peer-review experience. 
\revise{We focused on these two venues because they are top-tier venues in HCI and AI respectively, and have experienced substantial growth in submission volume.
CHI submissions grew more than fourfold from 1,520 in 2015 to 6,730 in 2026\footnote{\url{https://chi2026.acm.org/2025/12/12/insights-into-the-papers-track-first-round-outcomes/}}\footnote{\url{https://openaccept.org/c/chi/chi/}}, while AAAI submissions grew from 1,994 in 2015 to 23,680 in 2026\footnote{
\url{https://openaccept.org/c/ai/aaai/}}.}

To identify eligible participants, we manually collected author names and email addresses from the ACM Digital Library and the AAAI conference website. We sent invitation emails in batches of up to 500 recipients per day until we collected more than 200 valid responses. 
We used Mailmeteor\footnote{\url{https://mailmeteor.com/}} to schedule and automate the email distribution. 
The survey was designed and administered using Forms.app\footnote{\url{https://forms.app/}}.
Participants received approximately \$10 USD in compensation through Tremendous\footnote{\url{https://www.tremendous.com/}}, a global rewards platform that allows recipients to redeem rewards for regionally available gift cards. To ensure equitable compensation, we excluded participants from regions where the reward could not be redeemed through Tremendous.

For the CHI 2026 cohort, 5,504 authors were eligible after excluding authors from unsupported regions. We contacted 3,010 unique authors and received 128 completed responses (4.25\% completion rate); 120 passed the attention checks, and one additional response was excluded due to no peer-review experience, yielding 119 valid responses.

For the AAAI 2026 cohort, 3,445 authors were eligible after the same exclusion criteria. We contacted all eligible authors and received 98 completed responses (2.84\% completion rate); 85 passed the attention checks, and one additional response was excluded due to no peer-review experience, yielding 84 valid responses.
The detailed demographic information of survey study is listed in Appendix \ref{tab:survey demographics}.

\subsection{Survey Design}
We designed the survey to address \hyperref[RQ2.1]{\textbf{RQ2.1}} and \hyperref[RQ2.2]{\textbf{RQ2.2}}, drawing on findings from the previous interview study and relevant literature,  as described below (see Appendix~\ref{appendix: survey} for the complete survey questions). 

\subsubsection{Outcome Measures}
Drawing on findings from the interview study, we examined four outcome measures of reviewers' inline citation verification practices:
(1) \textbf{Perceived Importance.}
To capture how reviewers prioritize different types of inline citations (\revise{see Table~\ref{tab:citation_types} for the seven citation types}), participants rated the perceived importance of each citation type on a five-point Likert scale (i.e., \textit{How important are the following types of citations when assessing the quality of a manuscript during peer review?}).
(2) \textbf{Verification Coverage} and (3) \textbf{Verification Effort.}
To characterize reviewers' verification behavior, we measured both the coverage of citations they would verify and the effort required for verification. Participants reported these two measures separately using five-point response scales (i.e., \textit{When reviewing citations to unfamiliar work, how thoroughly do you typically verify each of the following 7
types?} and \textit{For each of the 7 types of inline citations, please indicate the extent to which you agree with the statement: ``I invest significant effort to ensure the correctness of arguments supported by this type of citation''}).
(4) \textbf{Desired AI Autonomy.}
To characterize reviewers' preferences for human-AI collaboration in inline citation verification, we adapted the classic Levels of Automation framework~\cite{sheridan1978human}. \revise{We condensed the original ten levels into five levels that capture distinct degrees of AI autonomy relevant to inline citation verification} (see Table~\ref{tab:autonomy-level}).

\begin{table}[t]
\centering
\caption{Desired AI autonomy levels for inline citation verification.}
\label{tab:autonomy-level}
\begin{tabular}{p{0.1\linewidth} p{0.8\linewidth}}
\toprule
\textbf{Level} & \textbf{Description} \\
\midrule
L1 & Handle the verification entirely by myself. \\
L2 & Use GenAI only for information retrieval (e.g., locating sources), while I perform the evaluation. \\
L3 & Use GenAI to flag potential issues, but make the final judgment myself. \\
L4 & Let GenAI conduct the initial verification and generate a report for my audit and refinement. \\
L5 & Let GenAI handle the verification entirely without my involvement. \\
\bottomrule
\end{tabular}
\end{table}

\subsubsection{Citation-Level Factor} 
Section~\ref{phase II: importance level} revealed that reviewers perceived varying levels of importance across different types of inline citations. 
To further examine whether these patterns generalize across citation practices, we grounded our survey design in an established citation taxonomy. 
Specifically, we adopted the seven inline citation categories proposed by Oppenheim et al.~\cite{oppenheim1978highly} and later refined by Bornmann et al.~\cite{bornmann2008citation} (see Table~\ref{tab:citation_types}). 
These categories enabled us to examine how participants' perceptions of importance and verification behaviors varied across different types of inline citations.

\subsubsection{Participant-Level Factors}
\paragraph{Demographic Factors}
We hypothesized that reviewers' research backgrounds and professional experiences would shape how they evaluate inline citations. Therefore, we collected demographic and background information, including peer-review experience, research field, and other relevant characteristics.
We expected that variations in reviewers' research field, reviewing experience, and regional contexts may lead to different evaluation practices and expectations regarding inline citation verification.

\begin{table}[t]
\centering
\caption{Seven inline citation types adopted in this study~\cite{oppenheim1978highly,bornmann2008citation}.}
\label{tab:citation_types}
\begin{tabular}{p{0.17\linewidth} p{0.73\linewidth}}
\toprule
\textbf{Citation Type} & \textbf{Description} \\
\midrule

Background &
Historical background (e.g. the author of the cited paper was the first person to work in the field; giving credit
for related work and providing background reading) \\
\hline
Relevant Work &
Description of other relevant work (e.g. a paper gives some relevant information; a paper makes a statement
with which the citing author agrees) \\
\hline
Theory &
Use of theoretical equation (i.e. the paper actually employed the theoretical equation cited for calculation
purposes) \\
\hline
Comparison &
Supplying information or data for comparison (i.e. a citing author made use of a cited article to obtain information
that was used for comparison) \\
\hline
Methodology &
Use of methodology (i.e. a citing author made use of a practical or theoretical technique given in the cited paper) \\
\hline
Critique &
A citing author felt it was necessary to state that a certain theory or method was not applicable or not the best one to use for their purpose, or a citing author criticized the cited paper \\
\hline
Other Data &
Supplying information or data, other than for comparison (i.e. a citing author made specific use of information
or data contained in a cited paper) \\

\bottomrule
\end{tabular}
\end{table}

\paragraph{Review Guideline Awareness}
Participants reported a lack of guidance on inline citation verification in official reviewer guidelines (see Section \ref{phase II: guideline}). 
In addition, we investigated whether reviewers carefully read the official guidelines and whether they noticed any guidance related to inline citation verification. 
To this end, we asked two questions: \textit{``How closely do you read reviewer guidelines before starting a review?''} and \textit{``Have you ever seen inline citation verification explicitly mentioned in reviewer guidelines?''}.

\paragraph{Self and GenAI Confidence}

Prior work by Lee et al.~\cite{lee2025the} showed that knowledge workers' self-confidence, confidence in GenAI, and confidence in evaluating GenAI outputs are associated with performance on critical thinking tasks. 
Given the interview participants' lack of confidence in GenAI (see Section \ref{Phase II: Confidence}), we included these three confidence measures to examine whether they are associated with reviewers' practices and perceptions regarding inline citation verification.

\paragraph{Trustworthiness}

Section~\ref{phase II: trustworthness} revealed that interview participants' trust in inline citations can change after being exposed to citation misuse. 
To investigate this factor, we adopted a measure inspired by prior work on source credibility and trustworthiness~\cite{hovland1951influence}. 
Specifically, we measured reviewers' trust in the trustworthiness of inline citations during the peer review process.

\subsubsection{Challenges and Improvements (Exploratory)}

Section~\ref{phase II: challenges} revealed that reviewers encounter a range of challenges when verifying inline citations.
To gain a deeper understanding of these challenges and identify opportunities for improving reviewer support, We included two questions: one multiple-choice question with an open-ended 'Other' option, \textit{``What challenges do you face when verifying inline citations?''}, and one open-ended question, \textit{``What tools would you like to have to help you verify inline citations during peer review?''}.

\subsection{Analysis}

\subsubsection{RQ2.1. Factors Associated with Perceptions and Practices}
We fitted cumulative link mixed models (CLMMs) using the \texttt{ordinal} package in \textsf{R} for each of the four output measures, treating responses as ordinal (1–5). 

The models included citation-level factor and participant-level factors as fixed effects. 
Citation type represented the citation-level factor, while participant-level factors included field, region, guideline awareness, confidence factors (confidence in self, in AI and in evaluating AI's output), perceived trustworthiness of inline citations, and review experience. 
We also included a participant-level random intercept to account for repeated responses across citation types, with \textit{Participant\_ID} defining the grouping structure.
We applied the Benjamini–Hochberg procedure~\cite{benjamini1995controlling} to adjust 
p-values for multiple comparisons within each outcome measures.

The statistical structure for each outcome measure (Perceived importance, Verification coverage, Verification effort and Desired AI autonomy) is specified as follows:

\begin{equation}
\begin{aligned}
\textit{Outcome Measure} \sim\;&
\text{Citation Type (7)}
+ \text{Guideline Awareness} \\
&+ \text{Confidence Factors (3)}
+ \text{Trustworthiness} \\
&+ \text{Review Experience}
+ \text{Field}
+ \text{Region} \\
&+ (1 \mid \text{Participant\_ID})
\end{aligned}
\end{equation}

For Desired AI autonomy, participants who indicated that they would not use GenAI in the question \textit{``Would you like to use GenAI, to any extent, to help assess whether the inline citations in a manuscript accurately
support the claims being made?''} were coded as having the lowest autonomy level  (i.e., 1) across all seven citation types.
Similarly, for the participant-level factor Review guideline awareness, participants who selected ``No'' in response to \textit{``Do you look at the reviewer guidelines before reviewing a paper?''} were coded as having the lowest level of guideline awareness (i.e., 1).

\subsubsection{RQ2.2. Challenges in Verifying Inline Citations and Desired Mitigation}

For the question regarding challenges \textit{``What challenges do you face when checking whether inline citations accurately support the claims made in the manuscript?''}, 13 participants provided additional open-ended responses beyond the predefined answer choices. Two authors coded these responses by mapping them to existing predefined categories, identifying emerging categories.

For the question regarding desired mitigation \textit{``What tools would you like to have to help you check whether inline citations accurately support the claims made in the manuscript?''}, two authors conducted an inductive coding process~\cite{thomas2006general}. They independently coded the responses, discussed disagreements, and iteratively refined the coding scheme until reaching consensus.

\begin{table*}[t]
\centering
\caption{Cumulative link mixed models (CLMMs) results for four measures of reviewers' inline citation verification practices: perceived importance, verification coverage, verification effort, and desired AI autonomy.}
\label{tab:clmm}
\small
\begin{tabularx}{.8\linewidth}{lRRRR}
\toprule

& Importance
& Coverage
& Effort
& AI Autonomy \\

\midrule

\multicolumn{5}{l}{\textbf{Model Information}} \\

Participants
& 203
& 203
& 203
& 203 \\

\midrule

\multicolumn{5}{l}{\textbf{Citation Type (Ref = Background)}} \\

Methodology
& \hlcell[75]{pos}{$1.56^{***}$}
& \hlcell[75]{pos}{$1.27^{***}$}
& \hlcell[75]{pos}{$1.44^{***}$}
& \hlcell[75]{neg}{$-0.60^{*}$} \\

Relevant Work
& \hlcell[75]{pos}{$1.31^{***}$}
& \hlcell[75]{pos}{$0.75^{***}$}
& \hlcell[75]{pos}{$1.09^{***}$}
& $-0.10$ \\

Comparison
& \hlcell[75]{pos}{$1.42^{***}$}
& \hlcell[75]{pos}{$1.06^{***}$}
& \hlcell[75]{pos}{$1.11^{***}$}
& $-0.37$ \\

Critique
& $0.30$
& $0.25$
& \hlcell[75]{pos}{$0.52^{*}$}
& \hlcell[75]{neg}{$-0.61^{*}$} \\

Theory
& \hlcell[75]{pos}{$0.85^{***}$}
& \hlcell[75]{pos}{$0.79^{***}$}
& \hlcell[75]{pos}{$1.03^{***}$}
& \hlcell[75]{neg}{$-0.75^{**}$} \\

Other Data
& \hlcell[75]{pos}{$0.44^{*}$}
& $-0.07$
& $0.30$
& $0.14$ \\

\midrule

\multicolumn{5}{l}{\textbf{Reviewer Characteristics}} \\

Guideline Awareness
& \hlcell[75]{pos}{$0.21^{*}$}
& \hlcell[75]{pos}{$0.39^{***}$}
& \hlcell[75]{pos}{$0.26^{*}$}
& $-0.12$ \\

Confidence in self
& $0.08$
& $0.17$
& $0.15$
& $-0.64$ \\
Confidence in AI
& $-0.17$
& $-0.06$
& $-0.11$
& \hlcell[75]{pos}{$1.17^{***}$} \\
Confidence in evaluation
& $0.17$
& $0.11$
& $0.06$
& $0.49$ \\

Trustworthiness
& \hlcell[75]{pos}{$0.39^{**}$}
& \hlcell[75]{pos}{$0.36^{*}$}
& $0.25$
& $-0.16$ \\

Review Experience
& $-0.01$
& $-0.01$
& $-0.03$
& $-0.07$ \\

\multicolumn{5}{l}{\textbf{Field (Ref = HCI)}} \\
\hspace{5mm}AI
& \hlcell[75]{pos}{$0.43^{*}$}
& $-0.05$
& \hlcell[75]{pos}{$0.71^{**}$}
& $-0.10$ \\
\hspace{5mm}Other
& $0.33$
& $-0.27$
& $-0.02$
& $1.67$ \\

\multicolumn{5}{l}{\textbf{Region (Ref = Europe)}} \\
\hspace{5mm}Americas
& $-0.27$
& $0.20$
& $0.10$
& $0.84$ \\
\hspace{5mm}Asia
& $0.23$
& $0.33$
& $0.20$
& \hlcell[75]{pos}{$2.64^{***}$} \\
\hspace{5mm}Oceania
& $0.46$
& $-0.02$
& $-0.33$
& \hlcell[75]{pos}{$3.08^{**}$} \\
\hspace{5mm}Other/Unknown
& $0.01$
& $0.93$
& $0.92$
& $2.54$ \\

\midrule

Random intercept variance
& 0.716
& 0.907
& 1.217
& 10.114 \\

\bottomrule
\end{tabularx}

\vspace{2mm}

\footnotesize{
Significance based on Benjamini--Hochberg adjusted $p$-values within each outcome:
$^{*}p_{\mathrm{adj}}<.05$,
$^{**}p_{\mathrm{adj}}<.01$,
$^{***}p_{\mathrm{adj}}<.001$.
}
\end{table*}

\begin{figure*}[t]
    \centering
    \includegraphics[width=\textwidth]{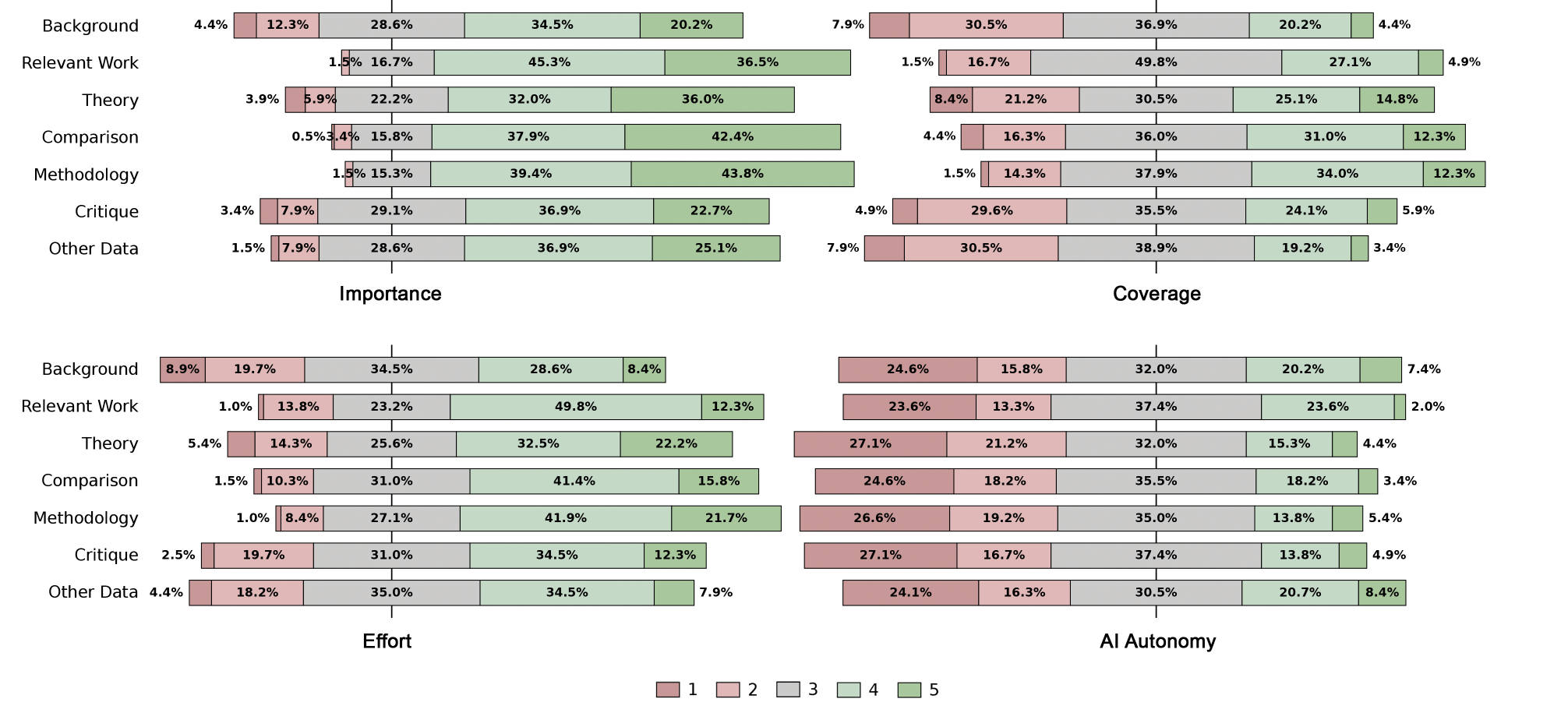}
    
    \caption{Distribution of responses across seven citation types for each output measure. 
    Importance was measured on a scale from 1 (\textit{Not at all important}) to 5 (\textit{Extremely important});
    Coverage from 1 (\textit{I do not verify them}) to 5 (\textit{I verify them in almost all cases}); 
    Effort from 1 (\textit{Strongly disagree}) to 5 (\textit{Strongly agree}) in response to the statement: \textit{``I invest significant effort to ensure the correctness of arguments supported by this type of citation''};
    AI Autonomy was measured using the five-level scale shown in Table~\ref{tab:autonomy-level}.}
    \Description{Distribution of responses across seven citation types for perceived importance, verification coverage, verification effort, and desired AI autonomy. Each stacked bar shows the percentage of participants selecting each response level from 1 to 5, with higher levels shown in progressively darker green. Percentages are displayed within each segment.}
    \label{fig:citation_types}
\end{figure*}

\subsection{Factors Associated with Perceptions and Practices  of Inline Citation Verification (RQ2.1)}
Table~\ref{tab:clmm} presents the CLMMs results. 
For each \revise{factor}, we report the fixed-effect coefficient $\beta$ on the cumulative logit scale with statistical significance indicated by asterisks.
The fixed-effect coefficient ($\beta$) represents the estimated change in the cumulative logit of the ordinal outcome associated with a one-unit increase in the predictor (or relative to the reference category for categorical predictors), holding all other variables constant. Positive (negative) coefficients indicate higher (lower) odds of responding in a higher category on the ordinal scale.

\subsubsection{Citation Types}
The distribution of responses across citation types for each \revise{measure} is shown in Figure~\ref{fig:citation_types}, while the corresponding CLMMs results are reported in Table~\ref{tab:clmm}, citation type was a significant predictor of reviewers' perceptions and verification practices. 
Compared with \textit{Background} citations, \textit{Methodology}, \textit{Relevant Work}, \textit{Comparison}, and \textit{Theory} citations were consistently perceived as more important, received broader verification coverage, and elicited greater verification effort (all $p_{\mathrm{adj}} < .001$). 
Among these, \textit{Methodology} citations showed the strongest effects (importance: $\beta = 1.56$; coverage: $\beta = 1.27$; effort: $\beta = 1.44$). 
In contrast, \textit{Critique} citations were associated only with greater verification effort ($\beta = 0.52$, $p_{\mathrm{adj}} < .05$), whereas \textit{Other Data} citations were associated only with higher perceived importance ($\beta = 0.44$, $p_{\mathrm{adj}} < .05$).

For desired AI autonomy, reviewers preferred lower levels of AI autonomy for \textit{Methodology} ($\beta = -0.60$, $p_{\mathrm{adj}} < .05$), \textit{Critique} ($\beta = -0.61$, $p_{\mathrm{adj}} < .05$), and \textit{Theory} ($\beta = -0.75$, $p_{\mathrm{adj}} < .01$) citations than for \textit{Background} citations.

\subsubsection{Reviewer Characteristics}
Greater review \textit{Guideline Awareness} was associated with higher perceived importance ($\beta = 0.21$, $p_{\mathrm{adj}} < .05$), broader verification coverage ($\beta = 0.39$, $p_{\mathrm{adj}} < .001$), and greater verification effort ($\beta = 0.26$, $p_{\mathrm{adj}} < .05$).
Out of 203 responses, 188 reviewers reported consulting reviewer guidelines, \textbf{more than half} of them (95/188, 50.5\%) indicated that the guidelines did not explicitly mention inline citation verification. 
Only a few reviewers (14/188, 7.5\%) reported that the guidelines provided detailed instructions on inline citation verification.
Others (79/188, 42.0\%) reported that they had only seen brief overview guidance on it.

Meanwhile, great \textit{Trustworthiness} of inline citations was positively associated with importance ($\beta = 0.39$, 
$p_{\mathrm{adj}} < .01$) and coverage ($\beta = 0.36$, 
$p_{\mathrm{adj}} < .05$).
Besides that, \textit{the confidence in AI} was positively associated with desired AI autonomy ($\beta = 1.17$, $p_{\mathrm{adj}} < .001$).

\subsubsection{Research Fields and Regions}
Compared with \textit{HCI} researchers, \textit{AI} researchers perceived inline citations as more important ($\beta = 0.43$, $p_{\mathrm{adj}} < .05$) and invested greater verification effort ($\beta = 0.71$, $p_{\mathrm{adj}} < .01$). 
In addition, compared with \textit{European} reviewers, reviewers from \textit{Asia} ($\beta = 2.64$, $p_{\mathrm{adj}} < .001$) and \textit{Oceania} ($\beta = 3.08$, $p_{\mathrm{adj}} < .01$) preferred higher levels of AI autonomy.

\subsection{\revise{Challenges and Strategies for Improving Inline Citation Verification (RQ2.2)}}

We summarize reviewers' reported challenges in Table~\ref{tab:challenges} and their desired support for inline citation verification in Table~\ref{tab:desired-tools}.

\subsubsection{Challenges}
As shown in Table~\ref{tab:challenges}, the most frequent challenge participants noted was time constraints (178/203, 87.7\%), suggesting that reviewers have limited time for citation verification amid increasing peer review demands. 
The review burden may further increase the mental effort required for citation verification, as nearly half of participants reported that verifying inline citations required substantial cognitive effort (96/203, 47.3\%).
Meanwhile, participants also highlighted practical barriers to tracing claims and their supporting evidence. 
Many struggled to locate the exact supporting passages (128/203, 63.1\%) within cited sources, while others faced difficulties accessing cited papers (82/203, 40.4\%), such as restrictions on library access. 
Moreover, domain unfamiliarity was another notable barrier (88/203, 43.3\%), which may make it more difficult for reviewers to assess whether cited evidence appropriately supports the claims being made.
Participants also reported a lack of transparency in the rationale behind inline citation choices (75/203, 36.9\%), making it difficult to determine why a particular citation was used to support a claim.

\subsubsection{Strategy: Citation Verification Support}

As shown in Table \ref{tab:desired-tools}, reviewers primarily sought support to help them identify and evaluate evidence, rather than fully automate the verification process. 
In particular, they wanted assistance in locating exact supporting passages in cited sources (72/203, 35.5\%) and assessing whether claims faithfully reflected their cited sources (38/203, 18.7\%).

When reviewing manuscripts outside their expertise or with insufficient background knowledge, participants also proposed support for expanding the context available for verification, such as help interpreting why a citation supports a claim (15/203, 7.4\%), retrieving additional context related to the cited source (5/203, 2.5\%), and providing additional domain knowledge to support verification (5/203, 2.5\%). 
Such support could help reviewers interpret cited evidence when their existing knowledge is insufficient to evaluate it.

\subsubsection{Strategy: User Interface Support}
Regarding how interfaces could reduce the effort involved in inline citation verification, participants emphasized cross-reference support (28/203, 13.8\%), such as displaying cited-source information in a pop-up card or providing easier access through a sidebar. These features could reduce the effort of switching between manuscripts and cited sources.
Moreover, reviewers preferred interfaces that highlight potentially problematic citations, such as weak support or citation mismatches (12/203, 5.9\%), and features that facilitate access to cited sources (8/203, 3.9\%) or visualize AI's confidence in assessing citation correctness (5/203, 2.5\%).
Together, these preferences suggest a desire for interfaces that reduce navigation effort and help reviewers prioritize citations requiring closer inspection.

\subsubsection{Adoption Considerations}
Our survey results revealed divergent preferences for adopting tools for inline citation verification. 
Although a considerable proportion of reviewers expressed a willingness to adopt AI-assisted tools (42/203, 20.7\%), others preferred manual verification (18/203, 8.9\%) or relying on search engines (18/203, 8.9\%). 
These differences reflected concerns about the reliability of AI (9/203, 4.4\%) and the potential for overreliance on AI to lead to deskilling among researchers (3/203, 1.5\%).

\begin{table*}[t]
\centering
\caption{Challenges reported when verifying whether inline citations support manuscript claims. Participants could select multiple options; percentages are within the analysis sample.}
\label{tab:challenges}
\begin{tabular}{lr}
\toprule
\textbf{Challenge} & \textbf{Participants} \\
\midrule
Time constraints & 178 (87.7\%) \\
Difficulty locating the exact statement & 128 (63.1\%) \\
\revise{Mental effort required for citation verification (Cognitive load)} & 96 (47.3\%) \\
Domain unfamiliarity & 88 (43.3\%) \\
\revise{Restricted access to cited papers} & 82 (40.4\%) \\
Unclear rationale behind authors' citation choices & 75 (36.9\%) \\
Lack of exposure to diverse perspectives & 16 (7.9\%) \\
Cross reference & 5 (2.5\%) \\
Language barrier & 1 (0.5\%) \\
\bottomrule
\end{tabular}
\end{table*}

\begin{table*}[t]
\centering
\caption{Themes in participants' desired support for inline citation verification. Responses could receive multiple codes.}
\label{tab:desired-tools}
\begin{tabular}{p{2.5cm}p{7.5cm}r}
\toprule
\textbf{Main theme} & \textbf{Sub-theme} & \textbf{Participants} \\
\midrule

\multirow{7}{2.5cm}{Citation\newline Verification\newline Support}
& Locate the exact supporting passage in the cited source & 72 (35.5\%) \\
& Verify that the claim faithfully reflects the cited source & 38 (18.7\%) \\
& Check citation exists (anti-hallucination) & 15 (7.4\%) \\
& Help interpret why the citation supports the claim & 15 (7.4\%) \\
& Summary or snippet of cited work for context & 14 (6.9\%) \\
& Retrieve additional context related to the cited source & 5 (2.5\%) \\
& Provide additional domain knowledge & 5 (2.5\%) \\

\midrule

\multirow{4}{2.5cm}{User\newline Interface\newline Support}
& Hover/click inline citation for quick reference lookup & 28 (13.8\%) \\
& Flags weak, unsupported, or mismatched citations & 12 (5.9\%) \\
& Easily access to the cited sources & 8 (3.9\%) \\
& Confidence score or color-coded verification status & 5 (2.5\%) \\

\midrule

\multirow{5}{2.5cm}{Adoption  Considerations}
& Prefer AI-assisted citation verification & 42 (20.7\%) \\
& Search engines (Google Scholar, etc.) & 18 (8.9\%) \\
& Prefer manual verification & 18 (8.9\%) \\
& Concern about AI reliability & 9 (4.4\%) \\
& Concern about AI-induced deskilling & 3 (1.5\%) \\

\bottomrule
\end{tabular}
\end{table*}

\section{Discussion}
\label{Section: Discussion}

\subsection{The Cross-Stakeholder Challenge of Inline Citation Verification}

\begin{figure}[t]
    \centering
    \includegraphics[width=.8\linewidth]{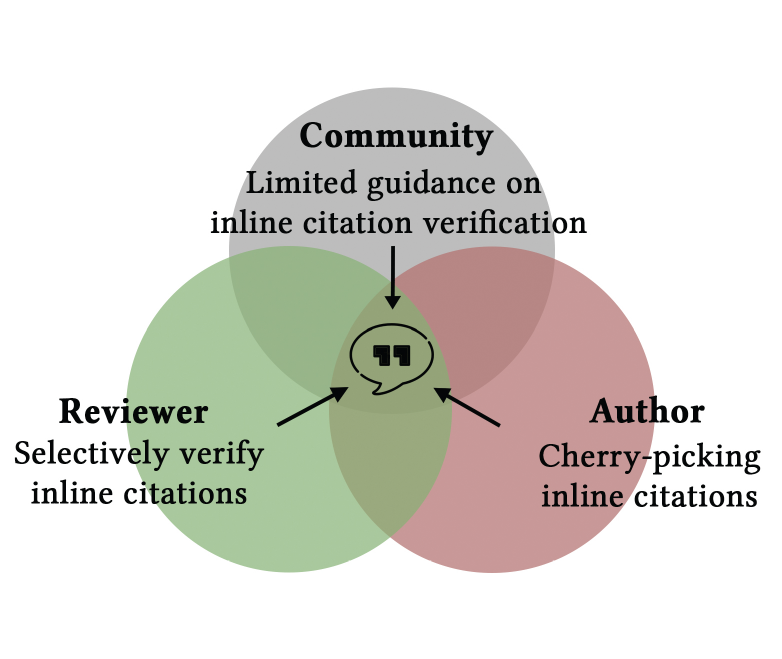}
    \caption{The cross-stakeholder challenge of inline citation verification: (1) authors may cherry-pick inline citations when producing manuscripts~\cite{fang2025exploring}; (2) reviewers selectively verify inline citations under peer-review; and (3) the research community lacks clear guidelines for verifying inline citations.}
    \Description{The figure illustrates the cross-stakeholder challenge of inline citation verification: (1) authors may selectively cite evidence when constructing manuscript claims; (2) reviewers may selectively verify inline citations under peer-review constraints; and (3) the research community lacks clear guidelines for inline citation verification.}
    \label{fig: cross_stakeholder_challenges}
\end{figure}

Inline citations are a fundamental mechanism for grounding scientific claims and establishing trust in scholarly communication. 
\revise{Yet our findings suggest that, although authors are ultimately responsible for the accuracy of their citations, the role of reviewers in verifying whether inline citations faithfully support the claims they accompany remains less clearly defined.}
Rather than representing isolated failures of authors or reviewers, problems surrounding inline citations emerge as a systemic issue in which responsibilities are fragmented across multiple stakeholders.

Situating our evidence within the broader literature on citation practices, we observed a consistent pattern of accountability gaps (see Figure~\ref{fig: cross_stakeholder_challenges}). 
Previous research~\cite{fang2025exploring} revealed authors may selectively choose evidence that supports their intended narratives, resulting in \emph{cherry-picking} citations that can misrepresent the underlying literature.
This issue is exacerbated by GenAI, as authors may uncritically incorporate hallucinated citations generated by GenAI into their papers~\cite{naddaf2026hallucinated,sakai2026hallucitation}, potentially reflecting confirmation bias~\cite{nickerson1998confirmation}.
Ideally, peer review should function as the primary safeguard against such practices. 
However, external pressures, such as limited time and increasing review workloads, constrain reviewers’ ability to systematically verify inline citations.
Rather than verifying citations comprehensively, we found reviewers implicitly prioritize citation checking based on the perceived importance of different citation types \revise{or personal interest}. 
This may leave unsupported or weakly supported claims undetected.

Our study further reveals that the current scholarly communication infrastructure provides limited support for establishing reviewer accountability. 
In our preliminary review guideline investigations, while the ICLR 2026 explicitly encourages reviewers to examine whether citations appropriately support the associated claims, most review guidelines provide little guidance regarding citation verification. 
Existing policies also tend to emphasize flexibility, for example, advising reviewers not to reject papers solely because of missing citations, without clarifying what constitutes unacceptable citation practices.
Our survey study also reveals that more than half of the participants who checked the review guidelines reported that they did not find any guidance on how to review inline citations.
As a result, reviewers are left to develop their own criteria for determining when citation issues warrant concern, leading to substantial inconsistencies in review practices.
 
This raises a fundamental question: \textbf{How can inline citation correctness be maintained through the collective efforts of different stakeholders?} 
We discuss design implications \revise{and considerations} on how to strengthen individual practices, facilitate cross-stakeholder collaboration, and foster transparent community norms for inline citation quality in the following sections. 
The summary of the following discussion is listed in Table \ref{tab:discussion takeaway}.


\begin{table*}[htp]
\centering
\small
\caption{Summary of discussion takeaways for inline citation verification in peer review.}
\setlength{\tabcolsep}{5pt}
\renewcommand{\arraystretch}{1.15}
\sloppy

\begin{tabularx}{\textwidth}{
    >{\raggedright\arraybackslash}p{2.5cm}
    >{\raggedright\arraybackslash}p{3.5cm}
    >{\raggedright\arraybackslash}X
}
\toprule
\textbf{Category} & \textbf{Subcategory} & \textbf{Takeaway} \\
\midrule

\multirow{17}{=}{%
    \raggedright
   
    \textbf{Design Implications \\ for Inline Citation Verification}\\

}
&
\textit{\autoref{sec:Reallocating} Reallocating Reviewer Verification Effort}
&
Citation verification support should reallocate, rather than increase, reviewers’ verification effort.
\\
\cmidrule(lr){2-3}

&
\textit{\autoref{sec:Adapting}
Adapting Verification Attention to Citation Types}
&
Verification attention should be calibrated to citation types, balancing human oversight and automation according to their criticality.
\\
\cmidrule(lr){2-3}

&
\textit{\autoref{sec:Augmenting}
Augmenting Reviewers’ Knowledge and Perspectives}
&
Citation verification should augment reviewers’ domain knowledge and perspectives to help them evaluate claims within broader scholarly context.
\\
\cmidrule(lr){2-3}

&
\textit{\autoref{sec:Designing}
Designing Explainable Inline Citations}
&
Inline citations should make authors’ evidence selection transparent by linking claims to specific supporting passages and providing optional rationales for citation choices.
\\
\cmidrule(lr){2-3}

&
\textit{\autoref{sec:Strengthening}
Strengthening Community Guidance Through Review Guidelines}
&
Review guidelines should establish clearer expectations for citation verification and embed these practices into reviewing workflows to support consistent reviewer accountability.
\\
\cmidrule(lr){2-3}

&
\textit{\autoref{sec:Learning}
Learning Across Research Communities}
&
Research communities should learn from one another to develop more consistent standards for inline citation verification while respecting disciplinary differences.
\\
\midrule

\multirow{6}{=}{%
    \raggedright
   
    \textbf{Considerations for Inline Citation Practices}\\

}
&
\textit{\autoref{sec:Preserving}
Turning Verification into Scholarly Learning}
&
Inline citation verification involves a tension between automation and opportunities for reviewers’ scholarly learning and engagement.
\\
\cmidrule(lr){2-3}
&
\textit{\autoref{sec:Preserving_Trust}
Preserving Trust in Citation Practices}
&
Inline citation practices involve a tension between AI-assisted scholarly writing and reviewers’ trust in supporting evidence.
\\
\cmidrule(lr){2-3}

&
\textit{\autoref{sec:Regional}
Regional Differences in Desired AI Autonomy}
&
AI-assisted citation verification practices may not be universally applicable across different socio-cultural contexts.
\\

\bottomrule
\end{tabularx}

\label{tab:discussion takeaway}

\end{table*}

\subsection{Design Implications for Inline Citation Verification}
\subsubsection{
\label{sec:Reallocating}
Reallocating Reviewer Verification Effort}
Our survey revealed a striking verification paradox. 
\revise{Although reviewers consistently rated inline citations as important, few reported verifying almost all unfamiliar inline citations} (see Figure \ref{fig:citation_types}, Importance and Coverage). 
The growing burden of peer review, driven by increasing submission volumes and citation counts~\cite{jonas2025past}, may make exhaustive verification unrealistic. 
Future inline citation review support should therefore aim to reallocate reviewers’ cognitive effort toward the most critical verification tasks, rather than introduce additional burdens~\cite{chen2026what}.
For example, as noted in Table \ref{tab:desired-tools}, future systems could serve as a collaborative verification partner by identifying potentially unsupported claims and citation mismatches, and prioritizing citations that warrant human attention ~\cite{lee2026reflens}.
In this model, future system would not replace citation verification, but help reviewers flag where verification effort is most worth investing. 
Such support would allow reviewers to reserve their limited cognitive resources for judgments that require human expertise.

\subsubsection{
Adapting Verification Attention to Citation Types}
\label{sec:Adapting}
Previous augmented interfaces for inline citation reading have helped readers cross-check references and connect citations with their reading history~\cite{citeSee2023}, but overlooked functional differences across citation types. Our results show that not all inline citations warrant the same degree of human oversight, with systematic differences across citation types in perceived importance, effort, verification coverage, and desired AI autonomy.
These findings suggest that future citation verification systems should move beyond a one-size-fits-all automation strategy.
Instead, AI autonomy could be calibrated according to citation criticality. 
Citation types, such as \emph{Background}, which reviewers perceived as less critical and more suitable for AI assistance, may benefit from higher levels of automation, including literature synthesis and evidence summarization. 
In contrast, citation types, such as \emph{Methodology}, which directly influence scientific validity, should remain under stronger human control. 

Lee et al.~\cite{lee2025the} found that GenAI can shift people's critical thinking toward tasks such as information verification and response integration.
Similarly, drawing on the cognitive processes defined in Bloom's taxonomy~\cite{anderson2001taxonomy}, one possible direction is to position AI primarily as a support for lower- to mid-level cognitive tasks, such as evidence retrieval, comprehension, and organization, while preserving reviewers' authority over higher-level cognitive tasks, including interpretation, evaluation, and final decision-making. 
Rather than maximizing automation, future systems should calibrate AI autonomy to both the cognitive demands of the task and reviewers' perceived AI preference. 
In practice, systems could allow reviewers to configure when AI should raise flags, recommend verification, or remain passive, making the level and timing of AI intervention responsive to both citation criticality and individual verification preferences. 
This suggests a shift from fixed automation toward \textbf{reviewer-controllable AI autonomy}, where reviewers retain agency over not only what the system does, but also when and to what extent it intervenes.

\subsubsection{
Augmenting Reviewers' Knowledge and Perspectives}
\label{sec:Augmenting}
A dominant challenge identified across our studies was domain unfamiliarity. Junior researchers often have less-developed domain expertise~\cite{choe2024supporting}, which can make it harder to interpret and evaluate scholarly claims. 
However, due to limited reviewer availability, junior researchers may still be asked to review submissions. Verifying whether a citation appropriately supports a claim often requires understanding the broader research landscape rather than reading the cited paper alone. 
Future reviewing systems should therefore augment reviewers' domain knowledge rather than merely retrieve citations. 
For example, systems could provide contextual explanations, summarize methodological traditions, and situate cited work within broader scientific conversations. 
Such support could help reviewers develop sufficient domain context to make informed judgments.

Our findings further suggest that citation verification is constrained not only by limited time but also by limited perspectives (see Table~ \ref{tab:challenges}). When reviewers approach a claim from a single disciplinary or conceptual perspective, they may be less likely to notice problematic citations or alternative interpretations. One possible direction is to use multi-agent systems to broaden the perspectives considered during citation verification, where agents with complementary disciplinary expertise engage in structured debates over whether cited evidence adequately supports manuscript claims. Such approaches have shown promise in critical paper reading~\cite{fang2026llm}, research ideation~\cite{liu2026perspectra}, and news reading~\cite{zhang2024see}. Rather than presenting a single recommendation, these systems could surface competing perspectives that encourage reviewers to critically reflect on their own reasoning.

\subsubsection{
Designing Explainable Inline Citations}
\label{sec:Designing}
Reviewers frequently reported two challenges during citation verification: \textit{Difficulty locating the exact statement} and \textit{Unclear rationale behind authors’ citation choices }.
These findings suggest that conventional inline citations provide insufficient information to support efficient and transparent verification.
Authors' intentions are difficult to observe directly~\cite{west2021misinformation}. 
Future citation systems should therefore support explainable inline citations.
Beyond linking to a reference, citations could identify the specific passage, figure, or result being cited and optionally for authors to provide a brief rationale for why the reference was selected. 
By making authors' evidence selection process more transparent, explainable citations can reduce reviewers' verification effort while strengthening accountability throughout the scholarly communication ecosystem.


\subsubsection{
Strengthening Community Guidance Through Review Guidelines}
\label{sec:Strengthening}
Our results showed that guideline awareness was positively associated with reviewers' perceived importance of inline citations, verification coverage, and verification effort (see Table~\ref{tab:clmm}). 
This finding highlights the role of review guidelines in shaping reviewers' inline citation verification practices, even when the guidelines provide vague or limited guidance on citation verification.

It suggests that reviewers may undervalue inline citation verification not necessarily because they lack motivation, but because their communities provide limited expectations regarding their responsibilities. 
As GenAI continues to reshape scholarly writing, clearer guidance is needed to define what constitutes sufficient citation verification and acceptable citation practices. 
Such guidance could also be embedded into in-situ reviewing interfaces (e.g., ReviewFlow~\cite{sun2024reviewflow}), enabling reviewers to verify citations within their existing workflow while reducing the additional cognitive burden.

\subsubsection{
Learning Across Research Communities}
\label{sec:Learning}
Our findings also revealed disciplinary differences in reviewers' perceptions of inline citations and their verification  between HCI and AI. 
These differences likely reflect not only disciplinary norms but also differences in reviewing cultures and community guidelines. 
For example, AI venues increasingly discuss \revise{attitudes toward citing arXiv papers}, whereas similar guidance remains less common in HCI venues.
Rather than developing reviewing practices in isolation, research communities should learn from one another when establishing reviewer guidelines. 
Cross-disciplinary collaboration may help identify effective practices and promote more consistent standards for inline citation verification across scholarly communities, ultimately strengthening the reliability of scientific communication.

\subsection{\revise{Considerations for Inline Citation Practices}}

\subsubsection{
Turning Verification into Scholarly Learning}
\label{sec:Preserving}

Automated inline citation misuse detection has been widely discussed in recent research~\cite{li2026bibagent} and can be used to check facts, but it remains unreliable~\cite{stokel2026ai}.
Similarly, our study also revealed that reviewers do not prefer to fully delegate inline citation verification to AI: fewer than 10\% of participants selected fully AI autonomy across seven types of inline citations (see Figure \ref{fig:citation_types}). 
Some of them also expressed concerns about AI's unreliability and the potential for over-reliance to undermine reviewers' verification skills (see Table \ref{tab:desired-tools}).

Importantly, reviewers described citation verification as an opportunity to keep up with recent developments and learn how researchers frame domain-specific contexts. 
Thus, citation verification may serve not only as a quality-control mechanism but also as an opportunity for scholarly learning and engagement. 
Our interview study also revealed that learning from the review process can motivate reviewers to engage in reviewing (see Section \ref{Interview-Growing}). 
This suggests a \textbf{tension between maximizing automation and preserving opportunities for scholarly learning.}
Future inline citation practices should therefore consider how automation may affect reviewers' engagement with cited evidence. 
For example, AI could surface relevant findings or related work during verification, while leaving reviewers to inspect and interpret the cited evidence themselves. 
Such learning opportunities may make the reviewing process more rewarding and motivate reviewers to engage more actively in verification.

\subsubsection{
Preserving Trust in Citation Practices}
\label{sec:Preserving_Trust}

Our survey revealed that reviewers' perceived trustworthiness was positively associated with both the perceived importance of inline citations and the effort invested in citation verification.
At the same time, recent work has highlighted the growing problem of AI-generated hallucinated citations~\cite{naddaf2026hallucinated}. 
As AI-generated citations become increasingly common, reviewers may become less confident in the reliability of supporting evidence and consequently less willing to invest effort in citation verification.

This suggests a \textbf{tension between AI-assisted scholarly writing and reviewers' trust in supporting evidence.}
Rather than asking reviewers to compensate for this uncertainty through additional checking, future scholarly communication systems should make citation provenance and verification status transparent. 
For example, systems could disclose AI usage~\cite{kusters2026more}, distinguish author-provided citations from AI-generated ones, indicate whether a citation has been independently verified against its source, and preserve an auditable record of how AI-generated citations were produced. 
This could allow reviewers to assess the reliability of citations at a glance, reducing uncertainty without requiring them to treat every citation as equally suspect.

\subsubsection{
Regional Differences in Desired AI Autonomy}
\label{sec:Regional}
We observed differences in researchers’ desired levels of AI autonomy for inline citation verification across geographic regions.
This observation is broadly consistent with prior cross-national research showing that perceptions of AI can vary across socio-cultural contexts.
For example, a survey of 24,790 participants across 20 countries found that national cultural dimensions, such as power distance and individualism, were associated with perceptions of AI~\cite{wang2025public}. 
Prior research has also reported regional differences in perceived usefulness of AI, with respondents in Korea reporting higher perceived usefulness than those in some other regions~\cite{kelley2021exciting}. 
These findings suggest that \textbf{AI-assisted citation verification practices may not be universally applicable across different socio-cultural contexts}.
Future inline citation practices should therefore remain sensitive to \revise{potential regional differences} in reviewers' preferences and expectations, rather than assuming a uniform level of AI autonomy.

\section{Limitations and Future Work}

First, our survey design adopted a commonly used seven citation-type categorization grounded in citation-function taxonomies from scientometrics~\cite{oppenheim1978highly,bornmann2008citation}.
However, these categories may not fully capture citation practices in research domains with different methodological traditions, such as thematic analysis and speculative design.
Future work should develop and validate domain-specific citation categories that better reflect disciplinary practices, enabling a more precise understanding of how researchers perceive and verify different types of inline citations.

Second, our analysis of publicly available review guidelines was intended as a preliminary exploration rather than a comprehensive comparison across venues.
We examined review guidelines from \revise{six} top-tier venues across HCI and AI to identify how venues frame reviewers' responsibilities toward citations and to motivate our subsequent empirical investigations.
As such, our analysis does not capture the full diversity of review guidelines across disciplines, venues, or publication cultures.
Future work should conduct a systematic and comprehensive analysis of review guidelines across a broader range of disciplines and venues to characterize how expectations for citation verification differ, identify common and effective practices, and reveal potential gaps or inconsistencies in existing guidelines.

Finally, the geographic distribution of participants in our survey was uneven, partly reflecting the distribution of researchers across CHI 2026 and AAAI 2026.
Some geographic groups were therefore represented by relatively small samples; 
for example, only 16 participants from Oceania responded to our survey, in general lower than regions such as the Americas and Europe (see Appendix \ref{tab:survey demographics}). 
In addition, our use of the predefined compensation platform, Tremendous, introduced a geographic limitation, as it did not support \$10 gift cards for recipients in China, resulting in the exclusion of participants from China.
Although our analysis considered field and demographic factors where appropriate, the uneven sample distribution limits the extent to which our findings can be generalized across geographic regions.
Future work should recruit more geographically balanced samples and examine whether citation perceptions, verification practices, and desired AI support vary across different research communities and regions.

\section{Conclusion}
We conducted a mixed-methods exploratory study to investigate how reviewers verify inline citations across HCI and AI fields. 
Our findings reveal that reviewers’ perceptions of citation importance, verification effort, and desired AI autonomy vary across different types of inline citations and are shaped by their awareness of review guidelines, trust in author-drafted citations, confidence in AI, and demographic region.
Taken together, our findings uncover two broader insights: (1) inline citations are not equally important to reviewers, leading them to allocate different levels of verification effort and seek different levels of AI support; 
and (2) citation accountability is distributed across multiple stakeholders in scholarly communication, yet remains fragmented in practice. 
By shifting citation verification from an individual reviewer responsibility toward shared accountability across the scholarly communication ecosystem, our work highlights opportunities for future citation verification support that promote more reliable and responsible use of inline citations.

\section*{GenAI Usage Disclosure}

We used GenAI at several stages of this research project. 
We used ChatGPT with its default model, GPT-5.6 Luna, to assist with refining the paper draft, survey questions and survey invitation text. 
We also used Cursor to assist with writing code for statistical analysis and generating Figure~\ref{fig:citation_types}.
In all cases, AI-generated outputs were reviewed and verified by the authors and revised to ensure their accuracy and appropriateness.

\begin{acks}
We thank all participants who took part in our study. This work was partly supported by the JST ASPIRE for Top Scientists (Grant Number JPMJAP2405), JST SPRING (Grant Number JPMJSP2108), and JSPS KAKENHI (Grant Number 26KJ0809).
\end{acks}

\bibliographystyle{ACM-Reference-Format}
\bibliography{reference}

\appendix
\section{Interview Script to Reviewers}
\label{intervew_scripit_reviewers}
\begin{enumerate}
    \item Could you briefly describe your most recent peer-review experience?

    \item How do you typically review inline citations during the peer-review process?

    \item What tools, methods, or strategies do you use to verify inline citations?

    \item When encountering citations that you are unfamiliar with, how do you determine whether they appropriately support the associated claims?

    \item Do you verify all inline citations? If not, what types of citations do you usually choose to examine or skip, and why?

    \item Under what circumstances do you provide feedback about inline citations in your review reports?

    \item Can you recall any examples where you commented on citation-related issues? What motivated you to do so?

    \item To what extent do you feel confident in judging whether inline citations are appropriate, accurate, and trustworthy?

    \item Are there citation-related issues that you believe reviewers may overlook or fail to detect? Why?

    \item Have you encountered citation integrity issues such as missing citations, selective citation practices, citation misrepresentation, or fabricated citations? If so, how did you identify them?

    \item How much attention do you typically devote to inline citations during peer review?

    \item What aspects of reviewing inline citations do you find cognitively demanding, tedious, or time-consuming?

    \item How do you think the effort required for reviewing inline citations has changed, or will change, over time?

    \item How important do you consider inline citations when evaluating a manuscript?

    \item How would you compare the importance of citation quality with other review criteria, such as novelty, methodological rigor, and clarity of writing?

    \item How would you assess the current reliability and trustworthiness of citation practices in academic publications?

    \item What factors do you think contribute to citation-related problems in academic papers?

    \item What kinds of support or tools would help reviewers evaluate inline citations more effectively?

    \item Which aspects of citation review do you think GenAI could assist with, and which aspects should remain under human judgment?

    \item How could authors make it easier for reviewers to evaluate the quality and appropriateness of inline citations?

    \item How do you envision the future of inline-citation review in the age of GenAI?
\end{enumerate}

\section{Interview Study Demographic Information}
\label{tab:pr_info}
\begin{table*}[h!]
\centering
\caption{Participant Demographics and Reviewing Experience}
\renewcommand{\arraystretch}{1.15}

\begin{tabular}{c c c c l l l}
\toprule
\textbf{ID} &
\textbf{Age} & 
\textbf{Gender} &
\textbf{Review Amount} &
\textbf{Review Role} &
\textbf{Current Role} &
\textbf{Major} \\
\midrule

PR1  & 22 & F & 6-10 & Area Chair (AC) & PhD Student & HCI \\
PR2  & 27 & M & 11-20 & External Reviewer & PhD Student & HCI \\
PR3  & 25 & M & 3-5 & External Reviewer & PhD Student & CV \\
PR4  & 29 & M & > 21 & Area Chair (AC) & Assistant Professor & ML \\
PR5  & 29 & M & > 21 & Associate Chair, PC Member & PhD Student & HCI \\
PR6  & 26 & M & 3-5 & External Reviewer & PhD Student & HCI \\
PR7  & 28 & M & > 21 & External Reviewer & Postdoctoral Researcher & CV \\
PR8  & 26 & M & > 21 & PC Member & PhD Student & ML \\
PR9  & 25 & M & 3-5 & External Reviewer & PhD Student & HCI \\
PR10 & 31 & M & 6-10 & External Reviewer & Assistant Professor & ML \\
PR11 & 39 & F &  > 21 & Associate Chair, PC Member & Assistant Professor & HCI \\
PR12 & 30 & F & 11 -20 &
\makecell[l]{Area Chair, Associate Chair,\\
PC Member, Associate Editor} &
Assistant Professor &
NLP \\
\bottomrule
\end{tabular}
\end{table*}

\section{Survey Questions}
\label{appendix: survey}

\begin{enumerate}[label=\arabic*., leftmargin=*]
    
     \item Enter your email address. [Free response]
    \item What is your gender?
    \begin{itemize}[label=$\square$, leftmargin=2em]
        \item Man
        \item Woman
        \item Non-binary / gender diverse
        \item Prefer not to say
    \end{itemize}

    \item How old are you? [Free response]
    
    \item What is your research field?
    \begin{itemize}[label=$\square$, leftmargin=2em]
        \item HCI (Human-Computer Interaction)
        \item AI (Artificial Intelligence)
        \item Other [Free response]
    \end{itemize}
    \item Please enter the country or region where your primary institution is located.[Free response]
    \item How many years of research experience do you have? (including PhD, industry research, etc.) [Free response]

    \item How many years of experience do you have in peer reviewing academic papers? [Free response]

    \item What roles have you taken in peer review? [Select all that apply]
    \begin{itemize}[label=$\square$, leftmargin=2em]
        \item External Reviewer
        \item PC Member
        \item Senior PC
        \item Area Chair
        \item Senior Area Chair
        \item Associate Chair
        \item Associate Editor
        \item No specific roles
        \item Other [Free response]
    \end{itemize}

    \item Do you look at the reviewer guidelines before reviewing a paper? (If No, skip to Question 12)
    \begin{itemize}[label=$\square$, leftmargin=2em]
        \item Yes
        \item No
    \end{itemize}

    \item How closely do you read reviewer guidelines before you start reviewing? (Scale from 1: \textit{Not at all closely} to 5: \textit{Extremely closely})

    \item Have you seen inline citation verification explicitly mentioned in guidelines?
    \begin{itemize}[label=$\square$, leftmargin=2em]
        \item Yes, with detailed instructions
        \item Yes, but only a brief overview
        \item No, not at all
    \end{itemize}

    \item What challenges do you face when checking whether inline citations accurately support the claims made in the manuscript? [Select all that apply]
    \begin{enumerate}[label=$\square$, leftmargin=2em]
        \item Time constraints
        \item Difficulty accessing cited papers
        \item Domain unfamiliarity
        \item Cognitive load
        \item Difficulty to locate the exact statement from long or complex reference source such as a book
        \item Lack of exposure to diverse perspectives during the review process
        \item Unclear rationale behind authors' citation choices
        \item Other [Free response]
    \end{enumerate}

    \item What tools would you like to have to help you check whether inline citations accurately support the claims made in the manuscript? [Free response]

    \item How confident are you in your ability to check whether inline citations accurately support the claims made in the manuscript without GenAI? 
    (Scale from 1: \textit{Not at all confident} to 5: \textit{Extremely confident})

    \item How confident are you in GenAI’s ability to check whether inline citations accurately support the claims made in the manuscript?
    (Scale from 1: \textit{Not at all confident} to 5: \textit{Extremely confident})

    \item How confident are you in your ability to evaluate GenAI generated outputs for checking whether inline citations accurately support the claims made in the manuscript?
    (Scale from 1: \textit{Not at all confident} to 5: \textit{Extremely confident})

    \item To confirm that you are paying attention, please select ``Extremely confident'' for this question. 
    (Scale from 1: \textit{Not at all confident} to 5: \textit{Extremely confident})

    \item How do you evaluate the trustworthiness of inline citations in the paper you recently reviewed? (e.g., whether the cited sources seem reliable, relevant, and correctly support the claims in the text)
    (Scale from 1: \textit{Not at all trustworthy} to 5: \textit{Extremely trustworthy})

    \item Please briefly explain the reasoning behind your trustworthiness rating. [Free response]

\begin{enumerate}[label=20-26., align=left, leftmargin=0pt] 
    \item How important are the following types of citations when assessing the quality of a manuscript during peer review? (Scale from 1: \textit{Not at all important} to 5: \textit{Extremely important}):
        \begin{enumerate}[label=(\alph*), leftmargin=*]
            \item Historical background (e.g. the author of the cited paper was the first person to work in the field; giving credit for related work and providing background reading)
            \item Description of other relevant work (e.g. a paper gives some relevant information; a paper makes a statement with which the citing author agrees)
            \item Use of theoretical equation (i.e. the paper actually employed the theoretical equation cited for calculation purposes)
            \item Supplying information or data for comparison (i.e. a citing author made use of a cited article to obtain information that was used for comparison)
            \item Use of methodology (i.e. a citing author made use of a practical or theoretical technique given in the cited paper)
            \item A citing author felt it was necessary to state that a certain theory or method was not applicable or not the best one to use for their purpose, or a citing author criticized the cited paper
            \item Supplying information or data, other than for comparison (i.e. a citing author made specific use of information or data contained in a cited paper)
        \end{enumerate}
\end{enumerate}

\begin{enumerate}[label=27-33., align=left, leftmargin=0pt] 
\item When reviewing citations to unfamiliar work, how thoroughly do you typically verify each of the following 7 types? (Scale from 1: \textit{I do not verify them} to 5: \textit{I verify them in almost all cases}):
    \begin{enumerate}[label=(\alph*), leftmargin=*]
        \item Historical background (e.g. the author of the cited paper was the first person to work in the field; giving credit for related work and providing background reading)
        \item Description of other relevant work (e.g. a paper gives some relevant information; a paper makes a statement with which the citing author agrees)
        \item Use of theoretical equation (i.e. the paper actually employed the theoretical equation cited for calculation purposes)
        \item Supplying information or data for comparison (i.e. a citing author made use of a cited article to obtain information that was used for comparison)
        \item Use of methodology (i.e. a citing author made use of a practical or theoretical technique given in the cited paper)
        \item A citing author felt it was necessary to state that a certain theory or method was not applicable or not the best one to use for their purpose, or a citing author criticized the cited paper
        \item Supplying information or data, other than for comparison (i.e. a citing author made specific use of information or data contained in a cited paper)
    \end{enumerate}
\end{enumerate}

\begin{enumerate}[label=34-41., align=left, leftmargin=0pt] 
    \item For each of the 7 types of inline citations, please indicate the extent to which you agree with the statement: \textit{``I invest significant effort to ensure the correctness of arguments supported by this type of citation.''} 
    (Scale from 1: \textit{Strongly disagree} to 5: \textit{Strongly agree})
    \begin{enumerate}[label=(\alph*), leftmargin=*]
        \item Historical background (e.g. the author of the cited paper was the first person to work in the field; giving credit for related work and providing background reading)
        \item Description of other relevant work (e.g. a paper gives some relevant information; a paper makes a statement with which the citing author agrees)
        \item Use of theoretical equation (i.e. the paper actually employed the theoretical equation cited for calculation purposes)
        \item To confirm that you are paying attention, please select "Strongly agree" for this question.
        \item Supplying information or data for comparison (i.e. a citing author made use of a cited article to obtain information that was used for comparison)
        \item Use of methodology (i.e. a citing author made use of a practical or theoretical technique given in the cited paper)
        \item A citing author felt it was necessary to state that a certain theory or method was not applicable or not the best one to use for their purpose, or a citing author criticized the cited paper
        \item Supplying information or data, other than for comparison (i.e. a citing author made specific use of information or data contained in a cited paper)
    \end{enumerate}
\end{enumerate}

    \item[42.] Would you like to use GenAI, to any extent, to help assess whether the inline citations in a manuscript accurately support the claims being made? (If No, skip to the end)
    \begin{itemize}[label=$\square$, leftmargin=2em]
        \item Yes
        \item No
    \end{itemize}

\begin{enumerate}[label=43-49., align=left, leftmargin=0pt] 
    \item For each of the following seven types of inline citation checks, please indicate the extent to which you would like to delegate this task to GenAI in verifying whether the citations accurately support the claims made in the manuscript.
    (Scale 1. \textit{Prefer to handle entirely by myself}; 2. \textit{Prefer GenAI assistance for information retrieval only (e.g., locating sources), while I perform the evaluation}; 3. \textit{Prefer GenAI to assist by flagging potential issues, but I make the final judgment}; 4. \textit{Prefer GenAI to conduct the initial verification and produce a report for my audit and refinement}; 5.\textit{Prefer GenAI to handle this task entirely without my involvement})
    \begin{enumerate}[label=(\alph*), leftmargin=*]
        \item Historical background (e.g. the author of the cited paper was the first person to work in the field; giving credit for related work and providing background reading)
        \item Description of other relevant work (e.g. a paper gives some relevant information; a paper makes a statement with which the citing author agrees)
        \item Use of theoretical equation (i.e. the paper actually employed the theoretical equation cited for calculation purposes)
        \item Supplying information or data for comparison (i.e. a citing author made use of a cited article to obtain information that was used for comparison)
        \item Use of methodology (i.e. a citing author made use of a practical or theoretical technique given in the cited paper)
        \item A citing author felt it was necessary to state that a certain theory or method was not applicable or not the best one to use for their purpose, or a citing author criticized the cited paper
        \item Supplying information or data, other than for comparison (i.e. a citing author made specific use of information or data contained in a cited paper)
    \end{enumerate}
\end{enumerate}
\end{enumerate}

\section{Survey Study Demographic Information}
\label{tab:survey demographics}
\begin{table*}[h!]
\centering
\caption{Participant demographics for the analysis sample, which has participants who failed the attention check removed ($N=203$). Continuous variables are reported as mean (SD). Categorical variables are $n$ (\% within column). Reviewer roles are multi-select and may sum to more than 100\%.}
\begin{tabular}{lrrr}
\toprule
 & CHI ($n=119$) & AAAI ($n=84$) & Total ($N=203$) \\
\midrule
\multicolumn{4}{l}{\textit{Continuous variables, $M$ (SD)}} \\
Age (years) & 33.90 (9.46) & 34.55 (9.09) & 34.17 (9.29) \\
Research experience (years) & 9.59 (7.87) & 10.37 (8.14) & 9.91 (7.97) \\
Review experience (years) & 7.36 (7.65) & 8.33 (7.88) & 7.76 (7.74) \\
\midrule
\multicolumn{4}{l}{\textit{Gender}} \\
Man & 69 (58.0\%) & 71 (84.5\%) & 140 (69.0\%) \\
Woman & 42 (35.3\%) & 10 (11.9\%) & 52 (25.6\%) \\
Non-binary / gender diverse & 4 (3.4\%) & 0 (0.0\%) & 4 (2.0\%) \\
Prefer not to say & 4 (3.4\%) & 3 (3.6\%) & 7 (3.4\%) \\
\midrule
\multicolumn{4}{l}{\textit{Primary research field}} \\
HCI & 94 (79.0\%) & 3 (3.6\%) & 97 (47.8\%) \\
AI & 9 (7.6\%) & 70 (83.3\%) & 79 (38.9\%) \\
Other & 16 (13.4\%) & 11 (13.1\%) & 27 (13.3\%) \\
\midrule
\multicolumn{4}{l}{\textit{Geographic region}} \\
Asia & 40 (33.6\%) & 14 (16.7\%) & 54 (26.6\%) \\
Europe & 37 (31.1\%) & 33 (39.3\%) & 70 (34.5\%) \\
Americas & 33 (27.7\%) & 27 (32.1\%) & 60 (29.6\%) \\
Oceania & 9 (7.6\%) & 7 (8.3\%) & 16 (7.9\%) \\
Other/Unknown & 0 (0.0\%) & 3 (3.6\%) & 3 (1.5\%) \\
\midrule
\multicolumn{4}{l}{\textit{Reviewer roles (multi-select)}} \\
External Reviewer & 111 (93.3\%) & 73 (86.9\%) & 184 (90.6\%) \\
PC Member & 56 (47.1\%) & 52 (61.9\%) & 108 (53.2\%) \\
Associate Chair & 41 (34.5\%) & 5 (6.0\%) & 46 (22.7\%) \\
Area Chair & 18 (15.1\%) & 18 (21.4\%) & 36 (17.7\%) \\
Senior PC & 15 (12.6\%) & 15 (17.9\%) & 30 (14.8\%) \\
Senior Area Chair & 8 (6.7\%) & 5 (6.0\%) & 13 (6.4\%) \\
Associate Editor & 23 (19.3\%) & 21 (25.0\%) & 44 (21.7\%) \\
Other& 17 (14.3\%)&9 (10.7\%) &26 (12.8\%) \\
\midrule
\multicolumn{4}{l}{\textit{Looks at reviewer guidelines}} \\
Yes & 110 (92.4\%) & 78 (92.9\%) & 188 (92.6\%) \\
No & 9 (7.6\%) & 6 (7.1\%) & 15 (7.4\%) \\
\midrule
\multicolumn{4}{l}{\textit{Wants to use GenAI for reviewing}} \\
Yes & 96 (80.7\%) & 63 (75.0\%) & 159 (78.3\%) \\
No & 23 (19.3\%) & 21 (25.0\%) & 44 (21.7\%) \\
\bottomrule
\end{tabular}
\end{table*}

\end{document}